\documentclass[11pt,a4paper]{article}
\usepackage{amsmath,amssymb,mathrsfs}
\usepackage[dvips]{graphicx}
\usepackage{epsfig}

\newcommand{\ms}[1]{\mathscr{#1}}
\newcommand{\mc}[1]{\mathcal{#1}}

\def\un#1{\relax\ifmmode\@@underline#1\else
        $\@@underline{\hbox{#1}}$\relax\fi}

\let\du=\du                     

\def\a{\alpha}

\def\c{\chi}
\def\d{\delta}

\def\f{\phi}

\def\j{\psi}

\def\l{\lambda}
\def\m{\mu}
\def\n{\nu}
\def\o{\omega}

\def\F{\Phi}

\def\L{\Lambda}

\def\ve{\varepsilon}

\def\car{{\cal R}}

\def\bo{{\raise-.3ex\hbox{\large$\Box$}}}               
\def\pa{\partial}                                       
\def\TH{{\raise.2ex\hbox{$\displaystyle \bigodot$}\mskip-4.7mu \llap H \;}}
\def\face{{\raise.2ex\hbox{$\displaystyle \bigodot$}\mskip-2.2mu \llap {$\ddot
        \smile$}}}                                      

\def\Bar#1{\overline{#1}}                       
\def\VEV#1{\left\langle #1\right\rangle}        
\def\leftrightarrowfill{$\mathsurround=0pt \mathord\leftarrow \mkern-6mu
        \cleaders\hbox{$\mkern-2mu \mathord- \mkern-2mu$}\hfill
        \mkern-6mu \mathord\rightarrow$}
\def\dvec#1{\vbox{\ialign{##\crcr
        \leftrightarrowfill\crcr\noalign{\kern-1pt\nointerlineskip}
        $\hfil\displaystyle{#1}\hfil$\crcr}}}           
\def\dt#1{{\buildrel {\hbox{\LARGE .}} \over {#1}}}     

\def\sfrac#1#2{{\vphantom1\smash{\lower.5ex\hbox{\small$#1$}}\over
        \vphantom1\smash{\raise.4ex\hbox{\small$#2$}}}} 
\def\bfrac#1#2{{\vphantom1\smash{\lower.5ex\hbox{$#1$}}\over
        \vphantom1\smash{\raise.3ex\hbox{$#2$}}}}       
\def\afrac#1#2{{\vphantom1\smash{\lower.5ex\hbox{$#1$}}\over#2}}    

\def\[{\lfloor{\hskip 0.35pt}\!\!\!\lceil}
\def\]{\rfloor{\hskip 0.35pt}\!\!\!\rceil}

\def\du#1#2{_{#1}{}^{#2}}
\def\ud#1#2{^{#1}{}_{#2}}

\def\ha{{\fracmm12}}
\def\tr{{\rm tr}}

\def\un{\underline}
\def\fracmm#1#2{{{#1}\over{#2}}}

\def\low#1{{\raise -3pt\hbox{${\hskip 0.75pt}\!_{#1}$}}}

\def\Dot#1{\buildrel{_{_{\hskip 0.01in}\bullet}}\over{#1}}
\def\dt#1{\Dot{#1}}

\newskip\humongous \humongous=0pt plus 1000pt minus 1000pt

\newif\ifdtup

\newcommand{\be}{\begin{equation}}
\newcommand{\ee}{\end{equation}}
\newcommand{\nbe}{\begin{equation*}}
\newcommand{\nee}{\end{equation*}}

\newcommand{\lb}{\label}

\def\lessim{\lower0.6ex\hbox{$\,$\vbox{\offinterlineskip\hbox{$<$}\vskip1pt\hbox{$\sim$}}$\,$}}
\def\grtsim{\lower0.6ex\hbox{$\,$\vbox{\offinterlineskip\hbox{$>$}\vskip1pt\hbox{$\sim$}}$\,$}}

\def\gor{{\mathfrak R}}

\def\caw{{\mathcal W}}

\def\caf{{\mathcal F}}
\def\gof{{\mathfrak F}}

\def\cae{{\mathcal E}}
\def\goe{{\mathfrak E}}

\begin{document}

\begin{titlepage}

\begin{center}

April 2014 \hfill IPMU14-0021\\

\noindent
\vskip2.0cm
{\huge \bf 

Starobinsky Model in N=2 Supergravity

}

\vglue.3in

{\large
Sergei V. Ketov~${}^{a,b,c}$ 
}

\vglue.1in

{\em
${}^a$~Department of Physics, Tokyo Metropolitan University \\
Minami-ohsawa 1-1, Hachioji-shi, Tokyo 192-0397, Japan \\
${}^b$~Kavli Institute for the Physics and Mathematics of the Universe (IPMU)
\\The University of Tokyo, Chiba 277-8568, Japan \\
${}^c$~Institute of Physics and Technology, Tomsk Polytechnic University\\
30 Lenin Ave., Tomsk 634050, Russian Federation \\
}

\vglue.1in
ketov@tmu.ac.jp

\end{center}

\vglue.3in

\begin{center}
{\Large\bf Abstract}
\end{center}
\vglue.1in

The Starobinsky inflationary model of $(R+R^2)$ gravity is extended to the new-minimal $(40+40)$ 
off-shell 4D, N=2 supergravity in N=2 chiral (curved) superspace. In its dual formulation, as a 
matter-coupled N=2 Einstein supergravity, inflaton belongs to a massive N=2 vector multiplet.

\end{titlepage}


\section{Introduction}\label{sec:Intro}

The simplest Starobinsky model \cite{star} of chaotic inflation is described by the higher-derivative gravity action 
(see Ref.~\cite{myrev} for a recent review) 
\be \lb {stara}
S[g] = \int \mathrm{d}^4x\sqrt{-g} \left[ -\ha R +\frac{1}{12M^2}R^2\right]
\ee
in terms of 4D spacetime metric $g_{\m\n}(x)$ having the scalar curvature $R$. We use the natural units
with the reduced Planck mass $M_{\rm Pl}=1$. During inflation in the high-curvature regime 
(with $H>> M$ and $|\dt{H}|<<H^2$), the second term in Eq.~(\ref{stara}) dominates over the  first one, and
the Starobinsky inflationary solution (attractor!) can be written down in the very simple form as
\be \lb{ssol}
H\approx \frac{M^2}{6}(t_{\rm exit}-t)~, \qquad 0<t\leq t_{\rm exit}~.
\ee
It describes a quasi-de-Sitter expansion of the early Universe and a Graceful Exit from inflation simultaneously. The model (\ref{stara}) has only one mass parameter $M$ that is fixed by the observational (CMB) data as $M=(3.0 \times 10^{-6})(\fracmm{50}{N_e})$ where $N_e$ is the e-foldings number, $N_e=(50\div 55)$. The predictions of the simplest Starobinsky model for the spectral indices  $n_s\approx 1-2/N_e\approx 0.964$, $r\approx 12/N^2_e\approx0.004$ and low non-Gaussianity are in agreement with the WMAP and PLANCK data ($r<0.13$ and $r<0.11$, respectively, at 95\% CL) \cite{planck2}, but are in {\it disagreement} with the recent  (March
2014) BICEP2 measurements ($r=0.2+0.07,-0.05$) \cite{bicep2}.

The action (\ref{stara}) can be dualized by the Legendre-Weyl transform \cite{lwtr}  to the standard quintessence acton (without higher derivatives, and without ghosts) with the celebrated scalar potential
\be \lb{starp}
V(\f) = \fracmm{3}{4} M^2\left( 1- e^{-\sqrt{\frac{2}{3}}\f }\right)^2
\ee
that is quite suitable for slow roll inflation at large positive $\f$ rolling down over the plateau.

The Starobinsky inflaton (dubbed {\it scalaron}) $\f$ of the mass M is thus identified with the spin-0 part of metric and has geometrical origin. This interpretation is obscure in the quintessence picture, but it is essential for fixing the inflaton interactions with other (matter) fields during reheating after inflation --- it is known as the universal reheating mechanism (in the original picture) \cite{myrev}.

It is, therefore, not surprising that all that has renewed interest in the Starobinsky model in the recent (since 2013) literature, especially in the context of N=1 supergravity \cite{yoko,ktt,el1,kl4,gre,el2,fklp1,ks3,fklp2,fkpro,kt2,
Ferrara:2013pla}, see also Refs.~\cite{gket,ket3,kl1,kstar,kl2,kl3,ktsu} for the related earlier work. 

The current status of the Starobinsky model is phenomenological because a quantized $(R+R^2)$ gravity is still
non-renormalizable and has the finite UV-cutoff given by the Planck mass  \cite{Hertzberg:2010,Kehagias:2013dr}.
The UV-completion of the Starobinsky model is to be sought in quantum gravity. In the context of superstring
theory, searching for the supergravity realization of the Starobinsky inflation is just the reasonable first step since
supergravity emerges as the low-energy effective theory of superstrings, with the Planck scale being the natural 
string scale.

In the context of non-supersymmetric models of inflation, there exist considerable freedom in reproducing
the PLANCK data \cite{Kehagias:2013dr}, whereas reproducing the BICEP2 data leads to the transplanckian
problem whose resolution requires the UV-completion. The  minimal 4D, N=1 supergravity relizations of the Starobinsky inflation  \cite{el1,kl4,gre,el2,fklp1,ks3,kt2} also have some (functional)  freedom in the choice of the effective theory (see Sec.~2).

It is, therefore, important to study compatibility of the Starobinsky inflationary model with supergravities beyond N=1 in 4D. This is the next natural step towards superstrings. Moreover, extended local supersymmetry severly restricts possible couplings, which is yet another argument for searching an extension of the Starobinsky model in extended supergravities. In this paper we get the affirmative answer in the case of the minimal 4D, N=2 supergravity.

Our paper is organized as follows. In Sec.~2 we briefly discuss the known realizations of the Starobinsky inflation in the 4D, N=1 old-minimal and new-minimal supergravities. In Sec.~2 we recall a geometrical construction of the (new) minimal 4d, N=2 supergravity in N=2 chiral curved superspace. Our new results about possible extensions of the Starobinsky model in N=2 supergravity are given in Sec.~4. Our conclusion is Sec.~5.

\section{Starobinsky model in N=1 supergravity}\label{sec:N1sugra}

The action (\ref{stara}) is a gravity theory with higher derivatives, it depends upon the scalar curvature $R$ only,
without its derivatives. We would like to preserve those properties in the minimal supergravity extensions of the model  (\ref{stara}). We are interested in a generic extension but with a minimal number of the fields involved. An
off-shell formulation of supergravity with auxiliary fields (and a closed algebra of gauge transformations) is most
suitable and convenient for those purposes because its supersymmetry transformations are independent upon an action.

The higher derivative N=1 supergravities in 4D can be constructed by using either the N=1 superconformal tensor calculus \cite{cec} or curved superspace \cite{ks3}. We employ here the superspace approach since it is manifestly supersymmetric and transparent. 

It is the common feature of all the higher derivative supergravity actions that some of the ``auxiliary" fields of the standard Einstein supergravity become dynamical. There are two standard minimal sets of the auxiliary fields in the Einstein supergravity, which are known as the {\it old-minimal} set~\cite{Ferrara:1978em,Stelle:1978ye,Fradkin:1978jq} and the {\it new-minimal} set~\cite{Sohnius:1981tp}, both having $12_{\rm B}+12_{\rm F}$ off-shell (and $2_{\rm B}+2_{\rm F}$ on-shell) field components. The old-minimal supergravity field representation was used for constructing the higher-derivative supergravity actions in Refs.~\cite{cec,ks3}, and its applications to the Starobinsky inflation were studied in Refs.~\cite{kl1,kl2,kl3,kl4,kt2}. The  new-minimal supergravity field representation was used for the same purposes in Refs.~\cite{gre,fklp1}. In the old-minimal approach (in the dual picture = the standard matter-coupled supergravity) the scalaron (inflaton) is identified with the real part of the leading (complex) scalar field component of a chiral (scalar) supermultiplet. In the new-minimal approach (in the dual picture) the inflaton (scalaron) is identified with the leading (real) scalar field component of a vector supermultiplet. In both the old- and the new- minimal N=1 supergravity extensions of the $(R+R^2)$ gravity one has $12_{\rm B}+12_{\rm F}$ off-shell and $6_{\rm B}+6_{\rm F}$ on-shell degrees of freedom, because some of the Einstein supergravity  ``auxiliary" fields become physical in the higher derivative supergravity. The new minimal extension of the higher-derivative supergravity is parameterized by a single real potential of the real scalar curvature superfield containing the Ricci scalar curvature $R$ as its $D$-type (last) field component \cite{fklp1}. The old minimal extension is  parameterized by a real potential of the two superfields, namely, the chiral one having the Ricci scalar curvature $R$ as its (last) $F$-type field component, and its conjugate. It should be emphasized that, unlike the Einstein supergravities, the old- and new- minimal field representations are not physically equivalent (on-shell) in the context of the higher derivative supergravities. 

The {\it linearized} $(R+R^2)$ supergravity was constructed by Ferrara, Grisaru and van Nieuwenhuizen in Ref.~\cite{fgn} where it was found that one needs the extra $4_{\rm B}+4_{\rm F}$ physical degrees of freedom indeed, beyond those present in the Einstein supergravity. Since the Starobinsky inflationary solution is essentially non-perturbative, a full non-linear extension of the $(R+R^2)$ gravity is required. A generic non-linear  N=1
supergravity extension of the $f(R)$ gravity actions was proposed by Cecotti in Ref.~\cite{cec}. Later we found 
in Ref.~\cite{kt2} that some of the actions of Ref.~\cite{cec} always have ghosts, namely, those having the $R$ dependence beyond the quadratic term. Hence, those actions should be ruled out, in agreement with similar
findings in Refs.~\cite{Ferrara:2013pla,Kaneda:2010qv}. The rest of the actions of Ref.~\cite{cec} reads
  in the curved superspace of the old-minimal N=1 supergravity as follows \cite{ks3,kt2}:
\begin{align} \label{NaF}
S_{N+F}= \int \mathrm{d}^4x\mathrm{d}^{4}\theta E^{-1}N(\mc{R},\bar{\mc R}) + \left[ \int \mathrm{d}^{4}x\mathrm{d}^{2}\Theta 2 \ms{E} 
F(\mc{R})+\text{H.c.}\right]
\end{align}
in terms of an arbitrary non-holomorphic real potential $N(\mc{R},\bar{\mc R})$ and an arbitrary holomorphic potential $F(\mc{R})$. Though the $F(\mc{R})$-dependence in the action (\ref{NaF}) can be absorbed (except of 
a constant) into the $N$-potential, we find more convenient to keep it here. 

Similarly to Ref.~\cite{kt2}, we use the standard (Wess-Bagger) notation in curved N=1 superspace  \cite{wb},
so that there is no need to repeat it here. The only essential difference is the oppostite use of Latin and mathcal letters
for denoting the scalar curvature $R$ and the corresponding $N=1$ scalar curvature superfield $\car$.

The bosonic part of the Lagrangian  $\mc{L}$ in Eq.~(\ref{NaF}) was derived in Ref.~\cite{kt2},
\begin{align} \label{gcomp}
e^{-1}\mc{L}_{\rm bos.}
& =\frac{1}{12}\left( 2N + 2N_{X}X+2N_{\bar{X}}X^{*}+F_{X}+\bar{F}_{\bar{X}}-8N_{X\bar{X}}X^{*}X-\frac{1}{9}N_{X\bar{X}}b^{a}b_{a}\right) R  \nonumber \\ 
& + \frac{1}{144}N_{X\bar{X}}R^{2} 
 -N_{X\bar{X}}\partial_{m}X^{*}\partial^{m}X  +\frac{1}{36}N_{X\bar{X}}\left( \ms{D}_{m}b^{m} \right)^{2} \nonumber \\
&-\frac{i}{3}b^{m}\left( N_{X}\partial_{m}X-N_{\bar{X}}\partial_{m}X^{*}\right) +\frac{i}{6}\ms{D}_{m}b^{m} \left( 2N_{X}X-2N_{\bar{X}}X^{*}+F_{X}-\bar{F}_{\bar{X}} \right) \nonumber \\
& -\frac{1}{18}\left( 2N+2N_{X}X+2N_{\bar{X}}X^{*}+F_{X}+\bar{F}_{\bar{X}}-8N_{X\bar{X}}X^{*}X-\frac{1}{18}N_{X\bar{X}}b^{b}b_{b}   \right) b^{a}b_{a}   \nonumber \\
&+16N_{X\bar{X}}(X^{*}X)^{2}+6FX^{*}+6\bar{F}X-4X^{*}X\left( -N+2N_{X}X+2N_{\bar{X}}X^{*}+F_{X}+\bar{F}_{\bar{X}}\right),
\end{align}
 where the subscripts denote the derivatives with respect to the given arguments.

One real d.o.f. associated with the vector field $b_a$ appears to be physical. However, it does not
contribute to slow-roll inflation, and cannot have a non-vanishing VEV \cite{kt2}. By those reasons, we ignore 
it in what follows. Then the relevant part of the Lagrangian (\ref{gcomp}) has the structure
\be \label{struc}
e^{-1}\mc{L}_{\rm bos.}= f_1(X,X^*)R + f_2(X,X^*)R^2 - f_{\rm kin.}(X,X^*)\partial_{m}X^{*}\partial^{m}X -V(X,X^*)~,
\ee
whose three structure functions $(f_1,f_2,f_{\rm kin.})$ of the complex scalar $X$ and its scalar potential $V$   
can be easily read off from Eq.~(\ref{gcomp}) in terms of the input potentials $(N,F)$. The scalar field $X$ 
generically acquires a non-vanishing $\VEV{X}=X_0$ that is determined by its scalar potential in vacuum,
\be \label{vac}
\frac{\pa V}{\pa X}=\frac{\pa V}{\pa X^*}=0~,
\ee 
 so that one arrives at the effective $(R+R^2)$ gravity model with
\be \label{effg}
e^{-1}\mc{L}_{\rm grav.}= f_1(X_0)R + f_2(X_0)R^2 -V(X_0)~,
\ee
whose coefficients can be related to those in Eq.~(\ref{stara}) by identifying
\be \label{dem}
f_1(X_0)=-\ha~,     \qquad f_2(X_0)=\frac{1}{12M^2}~~.
\ee 
A cosmological constant is supposed to play no role during inflation, so that the value of $V(X_0)$ should be
sufficiently small or zero. 

Of course, it implies certain (rather mild) restrictions on a choice of the potentials $(N,F)$. As a simple example, let
us consider the Ansatz \cite{kt2}
\begin{align} \label{ansatz}
F(X)=&F_{0}+F_{1}X \quad (f_{i} \in \mathbb{R})~,\\
N(X,X^*)=&n_{2}XX^*+\frac{n_{4}}{2}(XX^*)^{2} \quad (n_{i} \in \mathbb{R})~.
\end{align}
Then the physical sign of the kinetic term (no ghosts) , $f_{\rm kin.}>0$, requires
 \begin{align} \label{nogh}
n_2 + 2n_4|X_0|^2 >0~.
\end{align}
The same condition provides the physical sign to the $R^2$ term in Eq.~(\ref{gcomp}).

The scalar potential $V$ in our Ansatz (\ref{ansatz}) is 
bounded from below provided that
\begin{align} \label{n4sign}
n_4<0~.
\end{align}

The Einstein-Hilbert gravity term is recovered by demanding
\begin{align} \label{f1}
 F_1=-3~.
\end{align}
The vacuum equations  (\ref{vac}) have non-vanishing solutions for $X_0$ when $F_0\neq 0$. For sufficiently
small $F_0$, there is always a solution $X_0\approx  -\ha F_0$. The very similar models 
were considered in Refs.~\cite{kl1,kl2,kl3,kt2}, in order to stabilize the Starobinsky inflation in the old-minimal supergravity. As is clear from our discussion above, the $n_2$ should be positive, though it is not enough since
$n_4<0$. There are many ways of stabilization of the Starobinsky inflationary solution in the class of phenomenologically viable inflationary models defined by Eq.~(\ref{NaF}).
  
 A generic action reproducing $(R+R^2)$ gravity in the new-minimal supergravity approach reads \cite{fklp1}
\begin{align} \label{nj}
S_{J}= -3 \int \mathrm{d}^4x\mathrm{d}^{4}\theta E^{-1} \exp\left[-\frac{1}{3}J(V_R) \right] 
+ \frac{3\lambda}{2}\left[ \int \mathrm{d}^{4}x\mathrm{d}^{2}\Theta 2\ms{E} W^{\a}W_{\a}+\text{H.c.}\right]
\end{align}
in terms of an arbitrary real potential $J(V_R)$ of the real superfield $V_R$ whose N=1 vector superfield part
$V'_R$ has the scalar curvature term $-\frac{1}{6}R$ in its (last) $D$-type field component, with the standard
N=1 superfield strength $W_{\a}(V_R)=W_{\a}(V'_R)$.

Roughly speaking, any real scalar superfield $V$ is a sum of three irreducible superfields, the chiral one $\Lambda$,
the anti-chiral one $\Bar{\Lambda}$ and the vector one $V'$, as $V=\Lambda+ \Bar{\Lambda} +V'$. It is, therefore,
clear from Eq.~(\ref{nj}) that its first term describes a kinetic term of the chiral multiplet $\Lambda$ with
the K\"ahler potential $J(\Lambda+\Bar{\Lambda})$, coupled to supergravity (and with $R$, in particular) and having non-minimal couplings to the vector multiplet $V'$. The second term in Eq.~(\ref{nj})  has, in particular, the pure $R^2$ term. Therefore, the "scalar-tensor" part (relevant for inflation) of the theory (\ref{nj}) has the form
  \be \label{stnm}
\mc{L}_{\rm bos.}=  -\ha e^{-\frac{1}{3}J(\Lambda|+\Bar{\Lambda}|)}eR + \frac{\lambda}{12}eR^2 - eNLSM(\Lambda|,\Bar{\Lambda}|)~,
\ee
where the $NLSM$ stands for the Non-Linear Sigma-Model kinetic terms of the scalars $(\Lambda|,\Bar{\Lambda}|)$.
Since there is no scalar potential for them, there is no need for their stabilization. The  $R^2$ term alone is enough
for the Starobinsky inflation in the high curvature regime (see Sec.~1). The Einstein-Hilbert term can be canonically normalized by a Weyl rescaling of vierbein in Eq.~(\ref{stnm}). The Weyl rescaling does not affect the coefficient at the $R^2$ term and, hence, the scalaron mass $M^2 =\lambda^{-1}>0$ too, though it results in extra (derivative) couplings between $(\Lambda|,\Bar{\Lambda}|)$ and $R$. Hence, dynamics of the chiral superfield $\Lambda$ is not relevant for the  Starobinsky inflation in this picture. 

More details about the N=2 new-minimal (non-superconformal) extension of $(R+R^2)$ gravity can be found in Ref.~\cite{sub}.

In the dual picture to the new-minimal supergravity (as the matter-coupled standard supergravity) scalaron is identified with a real scalar $a$ of an N=1 massive vector  supermultiplet, and gets the scalar potential $V(a)= \frac{1}{2\l} (J_a)^2$, where $J_a=dJ/da$. \cite{gre,fklp1}. The physical sign of the kinetic term of $a$ requires $J_{aa}>0$ in our notation.

The origin of the effective supergravity theories (\ref{NaF}) and (\ref{nj}) should be sought in 4D compactified (closed) superstring theory, while all the supergravity couplings are expected to play a role {\it after} inflation, during reheating, because the scale of the supersymmetry breaking is expected to be much lower than the scale of the
Starobinsky inflation.

\section{N=2 chiral superspace supergravity setup}\label{sec:Setup}

The N-extended  {\it conformal} supergravity theory up to N=4 in 4D is most easily formulated as the gauge theory of the N-extended superconformal algebra. Then the unwanted ("truly superconformal") symmetries can be fixed by using the compensating N-extended "matter" supermultiplets --- the so-called  {\it compensators}. In the case of N=1 supergravity (Sec.~2), there are two (minimal) irreducible compensators given by a chiral N=1 multiplet and a tensor (linear) N=1 multiplet. Those N=1 compensators give rise to the old-minimal and the new-minimal field
representations of N=1 supergravity, respectively  \cite{wb}. 

As regards the N=2 case, one can employ either the N=2 superconformal tensor calculus \cite{rhvv,rvv}, the conventional N=2 curved superspace \cite{mbook}, or the N=2 harmonic superspace \cite{har}.  When hypermultiplets are excluded, a convenient, practical and geometrical description of N=2 supergravities with manifest N=2 supersymmetry is provided by the N=2 {\it chiral} curved superspace  with local $SO(2)$ symmetry \cite{m1,m2}. 
The notation adopted in  
Refs.~\cite{mbook,m1,m2} is based on the Wess-Bagger notation \cite{wb}, so that we refer the reader to 
Ref.~\cite{mbook} for a systematic introduction and the very explicit results about extended supergravities in
extended superspace. N=2 hypermultiplets in the context of the Starobinsky inflation will be considered elsewhere. 

An N=2 conformal (Weyl) supergravity multiplet has $24_{\rm B}+24_{\rm F}$ d.o.f. In order to descend to the N=2 Einstein supergravity, one needs {\it two} N=2 compensators, with one of them being an N=2  vector multiplet having $8_{\rm B}+8_{\rm F}$ off-shell  d.o.f.  The N=2 vector compensator breaks the super-Weyl
symmetry and the local $U(1)$ of the N=2 superconformal R-symmetry, but preserves its $SU(2)$ subgroup. The field content of an N=2 vector multiplet is given by 
\be \label{2vec}
\left( A,~~\l^A_{\a},~~a_m,~~N^{AB} \right)~,
\ee
where $A$ is a complex scalar, $\l^A_{\a}$ are two chiral spinors, $a_m$ is a real vector gauge field, and $N^{AB}$ is a symmetric tensor. The capital early Latin letters stand for internal indices, $A=1,2$, of N=2 supersymmetry.

There exist two minimal (called the old and the new) field representations of N=2 supergravity with 
$32_{\rm B}+32_{\rm F}$ off-shell d.o.f. We use the new-minimal off-shell irreducible N=2 field representation  \cite{m1} because it is most suitable for model building in N=2 chiral superspace. It breaks the local R-symmetry further, from the $SU(2)$ to its $SO(2)$ subgroup, and has the field content
\be \label{2mfr}
\left( e^a_m,~~\j_{mA}^{\a},~~B_m; \quad a_m,~~\l^A_{\a},~~C,~~N^{AB};\quad v_m,~~t'_{mn},~~M,~~b^A_{aB}\right)~,
\ee
where we have separated all fields into the three groups:  the fields of on-shell N=2 supergravity
$( e^a_m,~\j_{mA}^{\a},~B_m)$, the fields of the N=2 vector compensator $(a_m,~\l^A_{\a},~~C,~N^{AB})$, and
the four auxiliary fields. The vector field $v_m$ is the gauge field  of the $SO(2)$ symmetry.

However, it is not possible to construct a {\it local} N=2 invariant Lagrangian by using a minimal $(32+32)$ field representation only. There are several choices for the second N=2 compensator. The most natural one in N=2 chiral
superspace is given by an N=2 tensor (linear) multiplet with  $8_{\rm B}+8_{\rm F}$ off-shell  d.o.f.  It has the field content
\be \label{2ten}
\left( t_{mn},~~\c^A_{\a},~~C^{AB},~~F \right)~,
\ee
where $t_{mn}$ is an antisymmetric tensor, $\c^A_{\a}$ are two chiral spinors, $C^{AB}$ is the symmetric tensor,
and $F$ is a complex scalar. 

Altogether, it results in the minimal set of N=2 supergravity fields with $40_{\rm B}+40_{\rm F}$ off-shell d.o.f. 
\cite{mbook}. 

An N=2 vector multiplet was introduced by Grimm, Sohnius and Wess  \cite{gswess}.
The N=2 multiplets of linearized N=2 supergravity were obtained  by de Wit and van Holten \cite{vw}. An N=2 tensor multiplet in N=2 superspace was introduced by Gates and Siegel  \cite{gasi}.
The linearized N=2 Poincar\'e supergravity in terms of the N=2 irreducible superfields was  constructed in Ref.~\cite{keto}.

Full N=2 superspace $z^M=(x^m,\theta^A_{\a},\Bar{\theta}_A^{\dt{\a}})$   has the measure $\mathrm{d}^4x\mathrm{d}^8\theta$
of the vanishing (mass) dimension. It leads to the need of the N=2 superfield unconstrained pre-potentials, solving the N=2 supergravity constraints, for a construction of invariant actions.  More practical way is the use of N=2 {\it chiral\/} $U(1)$ superspace $(x,\Theta)$ that has only {\it four} $\Theta$'s. Indeed \cite{m2},
\begin{itemize} 
\item there exist an N=2 chiral curved superspace density $\goe$ with $\goe|=e$, similarly to N=1 chiral curved superspace (Sec.~2), where the vertical bar means taking all $\Theta$'s zero.
\item both N=2 vector and N=2 tensor (linear) multiplets can be described by the N=2 (restricted) chiral superfields,
$\caw$ and $\caf$, respectively,
\item the spacetime scalar curvature $R$ enters as a field component into one of the N=2 curvature superfields
$\gor$, which obeys the N=2 vector multiplet constraints after the gauge-fixing of the $SU(2)$ superconformal gauge
symmetry to $SO(2)$ by using the N=2 tensor multiplet compensator, as $(\frac{1}{4} \ms{D}^{\a A} \ms{D}_{\a}^A-8\Bar{\gor})\gor|=-\frac{1}{4}R+\ldots$, where the dots stand for the other ($R$-independent) terms.
\end{itemize}
Compared to Refs.~\cite{mbook, m1,m2}, we have changed the notation of N=2 superfields from $(\cae,R,W,\F)$ to
$(\goe,\gor,\caw,\caf)$, respectively, in order to avoid confusion with our previous Sections in this paper.

To be more specific, let us recall that the standard off-shell N=2 superspace constraints, defining the N=2 conformal 
supergravity with the $24_{\rm B}+24_{\rm F}$ field components,
can be solved in terms of a few (constrained) N=2 covariant (torsion and curvature) superfields which include the N=2 curvature superfield $\gor^{AB}$ that is symmetric with respect to its indices, has $U(1)$ weight 2, and contains the scalar curvature
$R$ amongst its field components --- see, e.g., Sec.~3 of Ref.~\cite{m1} for details.

The standard off-shell constraints defining the  N=2 Yang-Mills theory in N=2 curved superspace of N=2 conformal
supergravity are solved in terms of the N=2 chiral superfield strength $\caw$ having $U(1)$ weight 2 and obeying the conditions  \cite{mbook} 
\be \lb{chir}
\Bar{\ms{D}}^{\dt{\a}}_A\caw = \ms{D}_{\a}^A\Bar{\caw}=0
\ee
and
\be \lb{rch}
\left( \ha \[ \ms{D}^{\a A},\ms{D}_{\a}^B \]-8\Bar{\gor}^{AB}\right)\caw
+ \left( \ha \[ \Bar{\ms{D}}^A_{\dt{\a}},\Bar{\ms{D}}^{\dt{\a}B} \]-8\gor^{AB}\right)\Bar{\caw}=0~~,
\ee
defining the N=2 vector multiplet as the restricted chiral N=2 superfield with the field components (\ref{2vec}).
 The N=2 chiral Yang-Mills superfield strength $\caw$ is very similar to its N=1 superspace counterpart $W$  describing the N=1 Yang-Mills theory \cite{wb}. We need only an N=2 abelian version of $\caw$ for our purposes
in the next Sec.~4.

An N=2 tensor (linear) multiplet is described in curved N=2 superspace by a two-form gauge potential
$H=\ha \mathrm{d}z^m\wedge \mathrm{d}z^N H_{NM}$ subject to the gauge transformations 
$\d H=\mathrm{d}\o$ with the one-form gauge parameter $\o$. The N=2 tensor superfield strength three-form
$G=\mathrm{d}H$ is a hermitian superfield with {\it vanishing} $U(1)$ charge, and is subject to the N=2 constraints  \cite{gasi}. Like an N=2 vector multiplet, an N=2 tensor multiplet can also be represented as (restricted) part of an N=2 chiral superfield $\caf$ of $U(1)$ weight 2 as 
\be \lb{chit}
\Bar{\ms{D}}^{\dt{\a}}_A\caf = \ms{D}_{\a}^A\Bar{\caf}=0
\ee
and
\be \lb{rten}
G^{AB}=\frac{1}{4}\left( \ha \[ \ms{D}^{\a A},\ms{D}_{\a}^B \]-8\Bar{\gor}^{AB}\right)\caf
+ \frac{1}{4}\left( \ha \[ \Bar{\ms{D}}^A_{\dt{\a}},\Bar{\ms{D}}^{\dt{\a}B} \]-8\gor^{AB}\right)\Bar{\caf}~.
\ee
These conditions are invariant under the gauge transformations 
\be \lb{gtt}
\d\caf=\L~,
\ee 
whose parameter $\L$ is the N=2 restricted chiral superfield representing an N=2 vector multipet 
and obeying Eqs.~( \ref{chir}) and (\ref{rch}).

Therefore, unlike the N=1 case, an N=2 chiral superfield defines a  {\it reducible} representation of N=2 supersymmetry. An N=2 chiral multiplet can be decomposed into two irreducible (restricted chiral) N=2 multiplets, with one being an N=2 vector multiplet and another being an N=2 tensor multiplet --- see, e.g., 
Ref.~\cite{kbook} for details.

It is also possible to define an N=2 vector superfield $\caw_G$ out of an N=2 tensor superfield $G$ and its covariant derivatives in a highly non-linear way, such that $\caw_G$ obeys the constraints  (\ref{chir}) and (\ref{rch}). This construction is known in the literature as the N=2  {\it improved} tensor multiplet \cite{wro}. 

Since we did not use the second N=2 compensator yet, our considerations above equally
apply to N=2 vector and N=2 tensor multiplets in the N=2 conformal supergravity superspace with the $U(2)$ gauge symmetry too. Breaking the local $U(2)$ to the local $SO(2)$ in the (40+40) theory describing the N=2 superconformal  coupling of the N=2 minimal field representation (\ref{2mfr}) to the N=2 tensor compensator 
$\tilde{G}$ can be done by imposing the condition \cite{m1}
\be \lb{brte}
\tilde{G}\ud{A}{B}=-i\ve_{AB}~,
\ee
where $\ve_{AB}$ is the Levi-Civita symbol. In particular, the N=2 supergravity constraints together with the N=2
Bianchi identities imply that the gauge (\ref{brte}) yields
\be \lb{brr}
\gor_{AB}=\d_{AB}\gor~,
\ee
where the N=2 supergravity superfield $\gor$ satisfies the N=2 vector multiplet constraints  (\ref{chir}) and 
(\ref{rch}) --- seee Ref.~\cite{m1} for details. This is the key observation for constructing the N=2 supergravity actions (Sec.~4).

\section{Starobinsky model in N=2 chiral superspace supergravity}\label{sec:N1sugra}

Given the N=2 chiral density $\goe$, an N=2 vector (chiral) superfield $\caw$ and an N=2 tensor (chiral) 
superfield $\caf$, there exist only {\it two} basic constructions of invariant actions in N=2 chiral curved 
 superspace, namely, those of the N=2 Yang-Mills type with
\be \lb{ny1}
S_{\rm YM} \sim \int \mathrm{d}^{4}x\mathrm{d}^{4}\Theta \goe ~\tr\left( \caw\caw\right)+\text{H.c.}
\ee
and  those of the N=2 (abelian) vector-tensor type with
\be \lb{ny2}
S_{\rm VT} \sim \int \mathrm{d}^{4}x\mathrm{d}^{4}\Theta \goe ~\caf\caw+\text{H.c.}
\ee

For instance, the minimal N=2 supergravity $(40+40)$ action in N=2 chiral $U(1)$ superspace with the N=2
tensor compensator $\tilde{G}$ in the gauge (\ref{brte}) is given by \cite{m2}
\be \lb{res0}
S_{\rm SG} = -6 \int \mathrm{d}^{4}x\mathrm{d}^{4}\Theta \goe\tilde{\caf}\gor+\text{H.c.}
\ee
where the N=2 tensor (chiral) compensator superfield $\tilde{\caf}$ obeys the condition
\be \lb{rten2}
\frac{1}{4}\left( \ha \[ \ms{D}^{\a A},\ms{D}\low{\a B} \]-8\Bar{\gor}\ve\low{AB}\right)\tilde{\caf}
+ \frac{1}{4}\left( \ha \[ \Bar{\ms{D}}^A_{\dt{\a}},\Bar{\ms{D}}^{\dt{\a}}\low{B} \]-8\gor\ve\low{AB}\right)\Bar{\tilde{\caf}}
=\ve\low{AB}~.
\ee
in accordance to Eq.~(\ref{rten}). The action (\ref{res0}) is the {\it minimal} off-shell N=2 supersymmetric extension of the Einstein-Hilbert gravity action. On-shell it describes the coupling of the N=2 supergravity multiplet to an  abelian N=2 vector multiplet. The field component form of the action (\ref{res0}) was derived in Sec.~4 of Ref.~\cite{m2}. 

Our purpose in this Section is to find N=2 off-shell invariant actions in curved N=2 chiral $U(1)$ superspace, which extend $(R+R^2)$ gravity, without using extra N=2 matter superfields, i.e. only in terms of the minimal off-shell (40+40) field representaton of N=2 supergravity. Then the only actors we have for a construction of invariant 
N=2 superfield Lagrangians are given by the three N=2 chiral superfields:
\begin{itemize} 
\item the N=2 supergravity superfield $\gor$,
\item the N=2 abelian vector superfield $\caw_0$ of gravi-photon field $B_m$,
\item the N=2 tensor compensator $\tilde{\caf}$.
\end{itemize}
and the tools are provided by Eqs.~(\ref{ny1}) and (\ref{ny2}). The N=2 chirality and dimensional reasons forbid a 
 presence of the N=2 covariant derivatives in the N=2 chiral superfield Lagrangians. The gauge invariance of an N=2 chiral action requires that its N=2 chiral superfield Lagrangian has $U(1)$ weight +4. Moreover, the volume of the
N=2 chiral $U(1)$ superspace vanishes,
\be \lb{vol}
S_0 = \int \mathrm{d}^{4}x\mathrm{d}^{4}\Theta \goe =0~.
\ee
Therefore, the available resources for constucting chiral N=2 invariant actions are very limited.

Since the N=2 supergravity extension of the $R$ term is already available due to Eq.~(\ref{res0}), we just have to
find an N=2 supergravity extension of the $R^2$ term in terms of the N=2 superfields $\gor$ and $\caw_0$. It
is not difficult when using the tool (\ref{ny1}) together with the fact that the $R^2$ term is already generated from
the $\gor^2$ term after integration over $\Theta$'s. A generic N=2 invariant action reads
\be \lb{res1}
S_f = \int \mathrm{d}^{4}x\mathrm{d}^{4}\Theta \goe f(\gor, \caw_0)+\text{H.c.}
\ee
where $f$ is a {\it homogeneous} of degree 2 function of the N=2 superfields $\gor$ and $\caw_0$ (cf. the
construction of the N=2 invariant actions of N=2 vector multiplets coupled to N=2 supergravity in Ref.~\cite{rvv}).

A homogeneous of degree 2 function $f(x_1,x_2)$ obeys the differential equation
\be \lb{deq}
x_1\frac{\pa f}{\pa x_1} + x_2\frac{\pa f}{\pa x_2}=2f(x_1,x_2)
\ee
whose general solution is given by
\be \lb{deso}
f(x_1,x_2)=x_1x_2g\left(\frac{x_1}{x_2}\right)
\ee
in terms of arbitrary function $g(\frac{x_1}{x_2})$. In particular, the $f\propto x^2_1$ is reproduced by choosing
$g(\frac{x_1}{x_2})\propto \frac{x_1}{x_2}$. 

The desired off-shell N=2 supergravity extension of $(R+R^2)$ gravity is provided by a sum of Eqs.~(\ref{res0}) and (\ref{res1}). 

The natural question arises, as to whether it is possible to extend this action any further.  A possible extension is 
given by adding a linear term in $\caw_0$ to the N=2 supergravity action (\ref{res0}) as  \cite{m2}.
\be \lb{res2}
S_{\rm gSG} = -6 \int \mathrm{d}^{4}x\mathrm{d}^{4}\Theta \goe~\tilde{\caf}(\gor +g_0\caw_0)+\text{H.c.}
\ee
with a complex parameter $g_0$ of (mass) dimension 1. Adding this term leads to the gauged N=2 supergravity extension of $(R+R^2)$ gravity, with the gauge symmetry $SO(2) \times SO(2)$ and the action
\be \lb{res3}
S = -6 \int \mathrm{d}^{4}x\mathrm{d}^{4}\Theta \goe~\tilde{\caf}(\gor +g_0\caw_0)+
\int \mathrm{d}^{4}x\mathrm{d}^{4}\Theta \goe~ \gor~\caw_0~g\left(\frac{\gor}{\caw_0}\right)
+\text{H.c.}
\ee

The higher derivative $(R+R^2)$ gravity (like any phenomenologically relevant $f(R)$ gravity model as well) is known to be classically equivalent to the quintessence (or a scalar-tensor gravity) without higher derivatives --- see, e.g., Ref.~\cite{myrev} for details. Similarly, an N=1 supergravity extension of $f(R)$ gravity (Sec.~2) is known to be classically equivalent (or dual) to the standard matter-coupled N=1 supergravity with particular K\"ahler potential
and superpotential \cite{cec,ks3}.  It is instructive to understand, how does it work in the case of the N=2 supergravity defined by Eq.~(\ref{res3}).

Let us substitute the N=2 supergravity superfield $\gor$ in Eq.~(\ref{res3}) by an N=2 vector superfield
$\caw_s$ and simultaneously add to the action $S$ a new term,
\be \lb{lm}
S_{\gof} = \int \mathrm{d}^{4}x\mathrm{d}^{4}\Theta \goe~\gof (\caw_s -\gor)+\text{H.c.}~,
\ee
where we have introduced the N=2 chiral Lagrange multiplier superfield $\gof$. Varying the new action with respect to $\gof$ gives back the original theory (\ref{res3}).  The new (equivalent) full action is {\it linear}  in $\gor$, so that it
does not have higher derivatives. 

Roughly speaking, the N=2 chiral superfield $\gof$ is a sum of an N=2 vector multiplet $\caw_L$ and an N=2 tensor
multiplet $\gof_L$ as "$\gof_L=\caw_L+\gof_L$" (see Sec.~3). The term $\caw_L\caw_s$ in Eq.~(\ref{lm}) can be included into the $f$-term in Eq.~(\ref{res3}). Because of the gauge transformations (\ref{gtt}) of $\gof_L$, the $\caw_L$ can be absorbed into $\gof_L$ in the N=2 superfield factor present in front of the $\gor$ in Eq.~(\ref{lm}).
Therefore, it is convenient to choose that N=2 tensor multiplet  $\gof_L$ as the second N=2 compensator by using
the gauge condition  (\ref{brte}) on its N=2 field strength $G_L$ since it directly produces the $-\ha R$ term in the
corresponding (field-component) action.

The remaining extra terms in the full action, $(const. -6\tilde{\caf})\caw_s$, are {\it linear} in $\caw_s$ and do not have a kinetic term for $\tilde{\caf}$. Therefore, the constant  gives rise to a scalar potential
(and a mass term) for the N=2 vector multiplet  $\caw_s$ so that it becomes {\it massive} with  
$16_{\rm B}+16_{\rm F}$ off-shell  d.o.f.  The extra $8_{\rm B}+8_{\rm F}$ off-shell  d.o.f. are provided by the
N=2 tensor multiplet  $\tilde{\caf}$ that is "eaten up" by the N=2 vector multipet $\caw_s$.

Therefore, scalaron is one of the five scalars belonging to the massive on-shell N=2 vector multiplet 
in the dual (N=2 matter-coupled Einstein supergravity) description of the theory  (\ref{res3}). A derivation of the field-component actions is straightforward but requires a separate investigation.

\section{Conclusion}

We confirm the non-uniqueness of N=1 (non-conformal) extension of $(R+R^2)$ gravity and of the Starobinsky
model in the old-minimal and new-minimal off-shell formulations of N=1 supergravity in superspace, and
specify the existing freedom in terms of the relevant potentials and coupling constants, $(N,F_0)$ and $J$,
respectively. In the dual formulation of those higher-derivative supergravities as the standard N=1 (Einstein) 
matter-coupled supergravities, scalaron belongs either to a massive N=1 chiral (scalar) multiplet or a massive
N=1 vector multiplet, respectively, in agreement with Refs.~\cite{gre,el2,fklp1,kt2}.

Our main result is given by Eq.~(\ref{res3}). It describes the N=2 manifestly supersymmetric (off-shell) extension of
$(R+R^2)$ gravity with $40_{\rm B}+40_{\rm F}$ off-shell  d.o.f. in N=2 supergravity with the $SO(2) \times SO(2)$ gauged symmetry. The extension is parametrized by an arbitrary holomorphic potential $g$ and, in addition, has a complex dimensional coupling constant $g_0$. This extension is dual to the standard N=2 gauged supergravity 
coupled to the massive N=2 vector multiplet that has scalaron amongst its scalar field components.

The presence of a single holomorphic potential $g$ in the action (\ref{res3}) is remarkable because a generic N=4 conformal supergravity action is also parameterized by an arbitrary holomorphic function \cite{mbook}. Therefore, we expect that a similar N=4 supersymmetric extension of the $(R+R^2)$ gravity should exist (at least, with
on-shell N=4 supersymmetry) too.  Another
argument supporting this conjecture is the observation \cite{m2} that the minimal off-shell N=2 supergravity field representation (\ref{2mfr}) with $32_{\rm B}+32_{\rm F}$ off-shell  d.o.f.  is part of the N=4 supergravity theory of 
Ref.~\cite{Nicolai:1981t}.  Yet another supporting observation \cite{m2} is that the on-shell N=2 gauged supergravity theory (in its simplest form with $g=0$) is a truncation of the N=4 supergravity with the gauged $SU(2)\times SU(2)$ symmetry of Ref.~\cite{fsch}. 
 
It is an open question about a possibility of {\it partial} N=2 local supersymmetry breaking from N=2 to
N=1, that would allow us to directly connect the N=1 and N=2 supergravity theories considered in Secs.~2 and 4, respectively.

We use only {\it linear} N=2 multiplets in N=2 chiral superspace  of the new-minimal N=2 supergravity 
\cite{mbook,m1,m2}.  It may be possible to get similar results in the old-minimal $(40+40)$ off-shell N=2
supergravity with a nonlinear N=2 tensor multiplet as the second N=2 compensator \cite{Fradkin:1979jq}.

\section*{Note Added}

After a submission of the original version of our manuscript, a new paper \cite{Italy:2014} appeared,
where some specific models of the standard matter-coupled N=2 supergravity were studied in detail, in a search 
for an N=2 inflaton scalar potential. It is also worth mentioning that the N=2 extensions of the Starobinsky inflation  are likely to have a {\it higher} scale of inflation and a {\it larger} tensor-to-scalar ratio $r$ than that of the
simplest Starobinsky model (\ref{stara}).

\section*{Acknowledgements}

The author is grateful to S. Ferrara, R. Grimm and A. A. Starobinsky for discussions. This work was supported by the Tokyo Metropolitan University and the World Premier International Research Center Initiative (WPI Initiative), MEXT, Japan.

\end{document}